\documentclass[prb,aps,superscriptaddress,reprint]{revtex4-2}

\usepackage{graphicx}
\usepackage[colorlinks=true,urlcolor=blue,linkcolor=blue,citecolor=blue,]{hyperref}

\begin{document}

\author{Corinne Steiner}
\affiliation{JARA-FIT and 2nd Institute of Physics, RWTH Aachen University, 52074 Aachen, Germany}
\affiliation{Peter Gr\"unberg Institute  (PGI-9), Forschungszentrum J\"ulich, 52425 J\"ulich,~Germany}

\author{Rebecca Rahmel}
\affiliation{JARA-FIT and 2nd Institute of Physics, RWTH Aachen University, 52074 Aachen, Germany}

\author{Frank Volmer}
\affiliation{JARA-FIT and 2nd Institute of Physics, RWTH Aachen University, 52074 Aachen, Germany}
\affiliation{AMO GmbH, Advanced Microelectronic Center Aachen (AMICA), 52074, Aachen, Germany}

\author{Rika Windisch}
\affiliation{Institute for Theoretical Physics, TU Wien, 1040 Wien, Austria}

\author{Lars H. Janssen}
\affiliation{JARA-FIT and 2nd Institute of Physics, RWTH Aachen University, 52074 Aachen, Germany}

\author{Patricia Pesch}
\affiliation{JARA-FIT and 2nd Institute of Physics, RWTH Aachen University, 52074 Aachen, Germany}

\author{Kenji Watanabe}
\affiliation{Research Center for Electronic and Optical Materials, National Institute for Materials Science, 1-1 Namiki, Tsukuba 305-0044, Japan}

\author{Takashi Taniguchi}
\affiliation{Research Center for Materials Nanoarchitectonics, National Institute for Materials Science,  1-1 Namiki, Tsukuba 305-0044, Japan}

\author{Florian Libisch}
\affiliation{Institute for Theoretical Physics, TU Wien, 1040 Wien, Austria}

\author{Bernd Beschoten}
\affiliation{JARA-FIT and 2nd Institute of Physics, RWTH Aachen University, 52074 Aachen, Germany}

\author{Christoph Stampfer}
\affiliation{JARA-FIT and 2nd Institute of Physics, RWTH Aachen University, 52074 Aachen, Germany}
\affiliation{Peter Gr\"unberg Institute  (PGI-9), Forschungszentrum J\"ulich, 52425 J\"ulich,~Germany}

\author{Annika Kurzmann}
\affiliation{JARA-FIT and 2nd Institute of Physics, RWTH Aachen University, 52074 Aachen, Germany}
\affiliation{2nd Institute of Physics, University of Cologne, 50937 Köln, Germany}

\title{Current-induced brightening of vacancy-related emitters in hexagonal boron nitride}

\begin{abstract}
We perform photoluminescence measurements on vacancy-related emitters in hexagonal boron nitride (hBN) that are notorious for their low quantum yields. The gating of these emitters via few-layer graphene electrodes reveals a reproducible, gate-dependent brightening of the emitter, which coincides with a change in the direction of the simultaneously measured leakage current across the hBN layers. At the same time, we observe that the relative increase of the brightening effect scales linearly with the intensity of the excitation laser. Both observations can be explained in terms of a photo-assisted electroluminescence effect. Interestingly, emitters can also show the opposite behavior, i.e.~a decrease in emitter intensity that depends on the gate leakage current. We explain these two opposing behaviors with different concentrations of donor and acceptor states in the hBN and show that precise control of the doping of hBN is necessary to gain control over the brightness of vacancy-related emitters by electrical means. Our findings contribute to a deeper understanding of vacancy-related defect emitters in hBN that is necessary to make use of their potential in quantum information processing. 
\end{abstract} 

\maketitle
\textit{Introduction}
Hexagonal boron nitride (hBN) has emerged as a promising host material for quantum emitters~\cite{tran_quantum_2016, tran_quantum_2016-1, martinez_efficient_2016, choi_engineering_2016, caldwell_photonics_2019}. Especially vacancy-related defects, e.g.~defects consisting of a vacancy adjacent to a carbon substitutional, show significant potential for quantum information processing, as they can possess a triplet ground state and spin-conserved excited states~\cite{PhysRevMaterials.1.071001,sajid_defect_2018,CBVN-center-for-quibit,Ivady2020-qubit,review-spin-defects-in-hBN}. In this respect, spin-lattice relaxation times of defect spins in the microsecond range~\cite{spin-lattice-relaxation-microseconds-RT,Chejanovsky2021-defect-spin} and room-temperature optical initialization and readout of triplet defect states have already been demonstrated~\cite{Gottscholl2020}. However, whereas multi-carbon substitutional defects (e.g.~carbon dimers and trimers)~\cite{mendelson_identifying_2021, li_carbon_2022, kumar_localized_2023}, can result in emitters exhibiting high brightness even at room temperature~\cite{tran_robust_2016,chejanovsky_structural_2016, grosso_tunable_2017,jungwirth_temperature_2016,proscia_near-deterministic_2018, PR-1}, vacancy-related defects are notorious for their low quantum yields~\cite{PhysRevB.102.144105,review-spin-defects-in-hBN}. In fact, the low brightness of vacancy-related defects has so far prevented second-order correlation (g$^{(2)}$) measurements and thus the unambiguous confirmation of their single photon emission~\cite{review-spin-defects-in-hBN}. Therefore, methods for the brightening of such dark emitters are of high interest.
An effective method for tuning the emission energies and intensities of bright emitters was established by the integration of hBN into van der Waals heterostructures, which allow the use of optically transparent few-layer graphene gates~\cite{noh_stark_2018,yu_electrical_2022,white_electrical_2022}. This enables the precise control of the local electric field by applying voltages to the gates while maintaining optical access to the emitters. The reported variations in emitter intensity in response to the applied electric field have been attributed to several mechanisms, such as gate-induced (dis)charging of the emitter's charge transition levels or of nearby defects~\cite{noh_stark_2018, white_electrical_2022,yu_electrical_2022}.
These findings highlight the complex interplay between the local electronic environment and the optical properties of hBN emitters. A thorough understanding of this interplay might be crucial for paving the way towards quantum devices and sensors based on hBN-emitters~\cite{review-spin-defects-in-hBN,caldwell_photonics_2019,PR-1, fang_quantum_2024}.

\noindent Here we investigate the brightening of a dark hBN emitter that we assign to be a vacancy-carbon substitutional~(V-C) type defect by analyzing its phonon side bands (PSBs). Furthermore, we demonstrate that a gate-voltage induced leakage current across the hBN layers appears to be responsible for the brightening effect. 
Our findings can be explained by a model that includes photo-assisted electroluminescence as the underlying mechanism of the reported intensity variations. Additionally, our results enable us to assign the different gate-dependent intensity variations observed in different emitters to differences in the doping levels of the hBN. Therefore, our method of simultaneously measuring  gate-dependent photoluminescence (PL) and gate leakage currents provides an additional and useful method for the characterization of both the hBN host material and its emitters. 

%%%%%%%%%%%%%%%%% Figure 1 %%%%%%%%%%%%%%%%%%%%%%%
\begin{figure*}[]
\centering
\includegraphics[width=\linewidth]{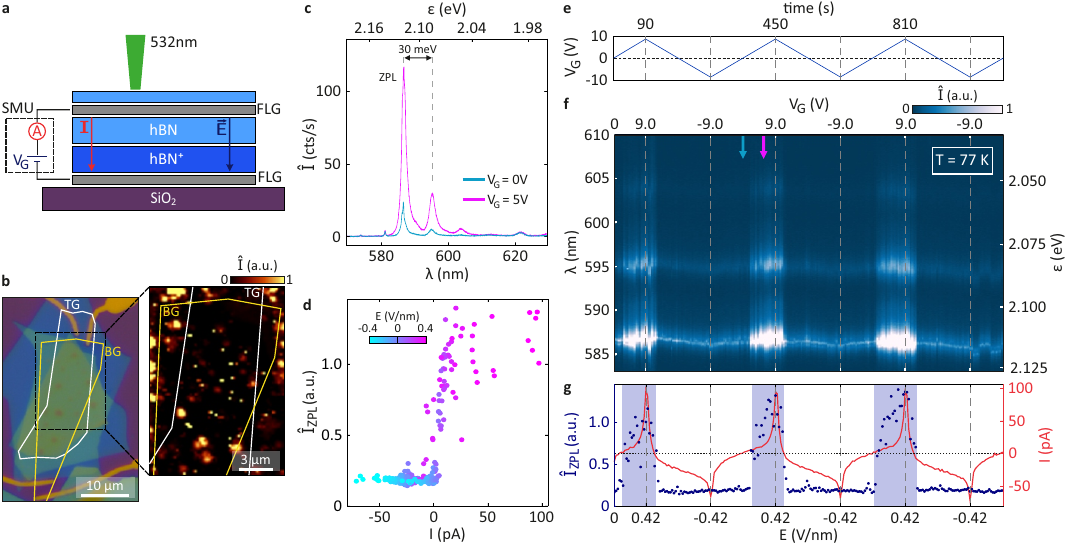}
\caption{
(a) Schematic cross-section of the few-layer graphene-gated hBN sample and its operating principle. The structure consists of an hBN emitter layer ($\mathrm{hBN^+}$) and an hBN capping layer that are encapsulated in few-layer graphene (FLG) and stacked onto a $\mathrm{Si/SiO_2}$ substrate. A source measurement unit (SMU) is used to apply a voltage ($V_\mathrm{G}$) to the FLG gates and to simultaneously measure the leakage current $I$ through the hBN.
(b) Optical image of the sample. The outlines of the top and back gate are highlighted with white and yellow lines, respectively. Zoom-in: Spatially-resolved PL signal of the dual-gated area. Bright localized spots indicate the observable emitters in the hBN. 
(c) Comparison of a high intensity emitter spectrum at $V_\mathrm{G} = 5$\,V, corresponding to $E = 0.23$\,V/nm, (magenta) and a low intensity spectrum at $V_\mathrm{G} = 0$\,V (cyan). The ZPL and two additional side modes are visible and show gate-dependent intensity variations. Both spectra were measured at 77\,K with a laser power of $\mathrm{200\,\mu W}$.
(d) The extracted zero phonon line (ZPL) intensity ($\hat{I}_\mathrm{ZPL}$) of one emitter scales with the measured leakage current (color coded to show its dependence on the applied external electric field $E$). 
(e) Illustration of how the gate voltage is swept in an alternating sequence over time between negative and positive values.
(f) Corresponding PL spectra of the emitter during repeated gate voltage sweeps measured at $T = 77$\,K and with a laser power of $\mathrm{200\,\mu W}$. The emitter intensity is plotted as function of the applied gate voltage and electric field, respectively. The dashed grey lines mark the turning points of the voltage sweeps. The emitter shows a reproducible linear Stark shift and voltage dependent changes in intensity. The blue and magenta arrows mark the spectra depicted in (c).
(g) ZPL intensity obtained by Lorentz fits on the ZPL peak from Fig.~\ref{f1}f (blue points). The blue shaded areas mark the electric field ranges where the emitter intensity is increased. The red curve shows the leakage current which was measured simultaneously to the PL signal. The dotted line indicates the zero-line of the current.}
\label{f1}
\end{figure*}
%%%%%%%%%%%%%%%%%%%%%%%%%%%%%%%%%%%%%%%%%%%%%%%%%%
\textit{Experimental methods}
The structure of the graphene-gated hBN samples used for this work is shown schematically in Fig.~\ref{f1}a. The samples consist of two hBN flakes which are encapsulated in few-layer graphene (FLG) flakes, resulting in a plate capacitor geometry. They were stacked and placed onto a $\mathrm{Si^{++}}$/$\mathrm{SiO_2}$ (285\,nm) substrate using a dry transfer method~\cite{Pizzocchero2016,Banszerus2017Feb,Bisswanger2022Jun}. Prior to the stacking, one hBN flake was thermally annealed (see Supplemental Material A~\cite{SM} for process details) to induce emitters into the hBN flake~\cite{tran_robust_2016} and is hereafter referred to as the emitter layer ($\mathrm{hBN^+}$). The second hBN flake acts as a capping layer, separating the surface of the emitter layer from the FLG in order to prevent quenching of surface emitters due to direct contact with graphene~\cite{stewart_quantum_2021,PhysRevLett.132.196902}. The FLG flakes serve as transparent, electrostatic gates, referred to as top gate (TG) and back gate (BG) and were contacted with lithographically defined Cr/Au contacts. Applying a gate voltage $V_\mathrm{G}$ between these two gates, i.e. electrodes, via a source measure unit (SMU, see Fig.~\ref{f1}a) allows to study emitters in hBN under the influence of the induced out-of-plane electric field. Additionally, we can probe the voltage-induced leakage current which flows between the two electrodes through the hBN layers. Fig.~\ref{f1}b shows an optical image of such a sample with yellow and white lines highlighting the outlines of the back and top gate, respectively. The zoom-in shown in the right panel of Fig.~\ref{f1}b depicts the spatially-resolved photoluminescence signal of the black-marked area. Localized emission centers are visible as spatially confined bright yellow spots.

\textit{Results}
We now focus on a specific emitter located within the dual-gated area of the sample, which exhibits a zero phonon line (ZPL) at approximately 587\,nm (2.11\,eV), see Fig.~\ref{f1}c. 
The first thing to note is the very low brightness of the emitter of only up to around 120 counts per second at an excitation power of 200\,$\mu$W and a laser spot size of about 500\,nm. Such a low quantum yield is a characteristic for vacancy-related defects~\cite{PhysRevB.102.144105,review-spin-defects-in-hBN}. 
Further support for the assignment of the emitter as a vacancy-related defect comes from the examination of the phonon side bands (PSBs) which can be observed in PL measurements conducted at a temperature of 77\,K (see Supplemental Material C~\cite{SM} for 4\,K data of the emitter). These PSBs are separated from the ZPL by an energy difference of around 30\,meV, which is much lower than the typical energy separation of $\mathrm{160\pm20\,meV}$ commonly observed for bright emitters~\cite{tran_quantum_2016,tran_robust_2016,martinez_efficient_2016,chejanovsky_structural_2016,mendelson_identifying_2021,fournier_position-controlled_2021,hayee_revealing_2020}. 
This reduced energy separation corresponds well with defect breathing modes of vacancy-related defects in hBN~\cite{tawfik_first-principles_2017, grosso_low-temperature_2020}. Due to the large spectral weight of these breathing modes~\cite{jara_first-principles_2021}, vacancy-related defects are not expected to show pronounced PSBs in the normally observable energy range. Moreover, the specific positions of the PSBs in the spectrum shown in Fig.~\ref{f1}c are found to be in agreement with theoretical predictions for a vacancy-carbon substitutional defect~\cite{tawfik_first-principles_2017,jara_first-principles_2021} suggesting that our experimentally observed emitter is indeed hosted by a V-C defect type (for further discussion, see Supplemental Material G~\cite{SM}).

\noindent Next, we focus on the gate voltage dependence of the emission properties of this dark emitter. As seen in Fig.~\ref{f1}c, the intensity of the emitter varies significantly between the two depicted spectra recorded at $V_\mathrm{G} = 0$\,V and $V_\mathrm{G} = 5$\,V. 
To investigate the brightening effect in more detail, we sweep the applied gate voltage $V_\mathrm{G}$ in repeated cycles between -9\,V and 9\,V, as illustrated in Fig.~\ref{f1}e. During these cycles, we simultaneously measure the leakage current flowing through the FLG/hBN/FLG structure and record PL spectra of the emitter. The color plot of Fig.~\ref{f1}f shows these spectra as a function of the resulting electrical field, which was calculated from the applied gate voltage according to the Lorentz local field approximation~\cite{noh_stark_2018, scavuzzo_electrically_2019} ($E$ = (($\epsilon_\mathrm{hBN}+2)\cdot V_\mathrm{G})/(3t)$, with the permittivity $\mathrm{\epsilon_{hBN} = 3.4}$ and the combined thickness $t=39$\,nm of both hBN layers). 
Over the repeated sweep cycles of the gate voltage we observe a linear Stark shift of $\mathrm{0.54 \pm 0.04\,nm/(V/nm)}$ and notably a clear and reproducible brightening of the emitter's ZPL and PSBs.
During the upward sweeps of the gate voltage, the emitter intensity increases abruptly before returning to its original intensity during the sweep back down to zero gate voltage. This is exemplary shown in Fig.~\ref{f1}c, which depicts two spectra at $V_\mathrm{G} = 0$\,V and $V_\mathrm{G} = 5$\,V that correspond to line-cuts in Fig.~\ref{f1}f (see blue and magenta arrows).
A direct comparison of the peak intensities, which were obtained by fitting a Lorentz curve to the measured spectra, reveals that the intensity increases by up to a factor of six. 
It is noteworthy that this increase only happens at positive gate voltages. Within the voltage range from $V_{\mathrm{G}} =$ 0\,V to -9\,V no increase in intensity is observed. Furthermore, by comparing the gate voltage values where the emitter switches its intensity from dark to bright with the values where the intensity switches back to its darker state it is apparent that the intensity variations follow a hysteretic behavior (see Supplemental Material H~\cite{SM} for further discussion).
In order to explore the interplay of the observed intensity variations with the leakage current through the hBN, the extracted peak intensities of the ZPL (blue points) are plotted alongside the simultaneously measured current (red curve) in Fig.~\ref{f1}g. This comparison reveals that the intensity variations of the emitter are directly linked to the current, $I$, through the hBN. Specifically, the transitions from negative to positive current directions (see the intersections between the red curve and the gray dashed line, the latter depicting the $I=0\,\mathrm{pA}$ level) coincide closely with the emitter switching between its dark and bright state. This observation becomes even more evident when the ZPL intensity is plotted as a function of the leakage current, as depicted in Fig.~\ref{f1}d. A distinct transition from low to high ZPL intensity is observed around zero leakage current, emphasizing the role of the current in modulating the brightness of the emitter. 
%%%%%%%%%%%%%%%%%% Figure 2 %%%%%%%%%%%%%%%%%%%%%%
\begin{figure}[]
\centering
\includegraphics[width=\linewidth]{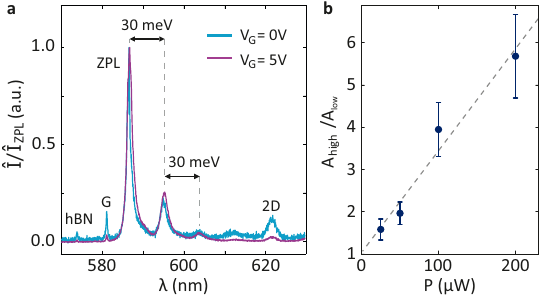}
\caption{
(a) Comparison of the two spectra shown in Fig.~\ref{f1} c with their intensity normalized to 1. For the sake of clarity, peaks related to the Raman signals of hBN (574\,nm~\cite{li_temperature_2018}) and graphene (G-peak at 581\,nm and 2D peak at 620\,nm~\cite{ferrari_raman_2006}) which do not scale have been labeled. 
(b) Ratio between the bright and dark ZPL amplitudes as a function of the power of the excitation laser (laser spot size of around 500\,nm). The dashed grey line serves as a guide to the eye and depicts ideal linear behavior.  
}
\label{f2}
\end{figure}
%%%%%%%%%%%%%%%%%%%%%%%%%%%%%%%%%%%%%%%%%%%%%%%%%%
We note that, except for the shift in wavelength due to the Stark effect, the ZPLs as well as the PSBs match each other for both the dark and the brightened states, as is evident when comparing spectra of the emitter with the ZPL normalized to 1 as shown in Fig.~\ref{f2}a.
Furthermore, we identify that a laser-induced photo-excitation process must play a crucial role in the brightening effect of the emitter distinguishing it from a pure electroluminescence effect \cite{park_narrowband_2024, yu_electrically_2025}. For this, Fig.~\ref{f2}b shows the ratio of the ZPL amplitudes in the bright and dark states as a function of laser excitation power, which reveals that the brightening effect becomes increasingly pronounced towards higher laser powers. We are not aware of any electrostatic (dis)charging effects, which were previously attributed to intensity variations of bright hBN emitters~\cite{noh_stark_2018,yu_electrical_2022,white_electrical_2022}, that would depend on laser power in such a way. It can thus be concluded that the brightening of the dark emitter necessitates a photo-excitation process that is also closely linked to the leakage current across the hBN layers, as shown in Fig.~\ref{f1}d.

\begin{figure*}[]
\centering
\includegraphics[width=\linewidth]{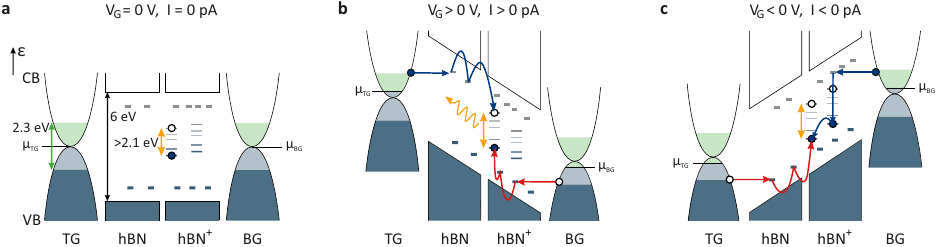}
\caption{
Model that links changes in the charge transport across the hBN layers to the intensity variations of the emitter in the hBN$^+$ layer. The top gate (TG) and the bottom gate (BG) are made of few-layer graphene. 
(a) At $V_\mathrm{G} = 0$\,V no charge transport through the hBN occurs and the emitter can only be excited by optical absorption in the hBN$^+$ layer.
(b) For $V_\mathrm{G} > 0$, photo-excited electrons (filled circles) and holes (open circles) from the gate electrodes, (shaded areas in the band structure of the FLG), undergo a photo-assisted field emission into carbon-related donor and acceptor states, depicted by blue and red arrows, respectively. Electrons and holes flow towards the emitter in the hBN$+$ via laser-induced Poole-Frenkel like emission (blue and red curved lines) and recombine over the energy levels of the emitter, creating the brightening effect due to electroluminescence.
(c) For $V_\mathrm{G} < 0$, electrons injected by the back gate electrode scatter into mid-gap states of the emitter layer before reaching the emitter, suppressing the electroluminescence effect.}
\label{f3}
\end{figure*}

\noindent Based on the above arguments, we present a model (see Fig.~\ref{f3}) that proposes a photo-assisted injection of charge carriers - driven by the interplay of laser excitation and gate voltage - as the mechanism responsible for the brightening of the emitter. 
%
%%%%%%%%%%%%%%%%%% Figure 3 %%%%%%%%%%%%%%%%%%%%%%
%%%%%%%%%%%%%%%%%%%%%%%%%%%%%%%%%%%%%%%%%%%%%%%%%%%%%%%%%%
We approximate the gapless band structure of the FLG top and bottom gates (TG and BG) by two parabolic bands~\cite{PhysRevLett.104.176404}. At zero gate voltage, the respective chemical potentials $\mu_{\mathrm{TG,BG}}$ of the two gates are positioned at the charge neutrality points (Fig.~\ref{f3}a). 
Under laser illumination ($E=2.3$\,eV), electrons in the gate electrodes are excited from the valence into the conduction band (indicated by the blue and green shaded areas in Fig.~\ref{f3}a), creating photo-excited electrons and holes up to 1.15\,eV away from the chemical potentials. We note that scattering processes among these photo-excited charge carriers may push a small fraction of them to even higher energy states~\cite{laser-induced-hot-carriers-1,laser-induced-hot-carriers-2,laser-induced-hot-carriers-3}.
We now discuss the mechanism by which charge carriers can be injected from the FLG electrodes into the hBN layers. First, we note that the electric field strength at which the leakage current across the hBN begins to show a strongly non-linear increase in Fig.~\ref{f1}g is rather small in comparison to reference samples that are not showing a current-induced brightening effect (see Supplemental Material F~\cite{SM}). We attribute this low threshold voltage to the onset of photo-assisted field emission~\cite{photo-assisted-field-emission-1,photo-assisted-field-emission-2} into carbon monomer defect ($\mathrm{C_B}$, $\mathrm{C_N}$) states present in the hBN~\cite{leakage-due-to-carbon-defects,defect-tunneling} (see Supplemental Material F~\cite{SM} for detailed discussion). These defects can form both donor and acceptor states based on whether the carbon atom replaces a boron ($\mathrm{C_B}$) or a nitrogen atom ($\mathrm{C_N}$)~\cite{energy-levels-2,energy-levels-3,energy-levels-7,energy-levels-6,mackoit-sinkeviciene_carbon_2019,carbon-dimer-energy-levels-1}, as illustrated by the discrete energy states (grey and blue short dashes) in the hBN band gap in Fig.~\ref{f3}. Depending on the calculation method~\cite{DFT-problematic}, the corresponding energy states are predicted to lie anywhere from a few hundred meV up to 1.5\,eV below the conduction band or above the valence band edge~\cite{energy-levels-2,energy-levels-3,energy-levels-7,energy-levels-6,mackoit-sinkeviciene_carbon_2019,carbon-dimer-energy-levels-1}. Furthermore, carbon monomers have been proposed to conduct gate leakage currents across the hBN layers~\cite{leakage-due-to-carbon-defects,defect-tunneling}. 
We therefore assume that under the application of a gate voltage (see Fig.~\ref{f3}b), photo-excited charge carriers from the gate electrodes undergo photo-assisted field emission~\cite{photo-assisted-field-emission-1,photo-assisted-field-emission-2} into these carbon monomer states (depicted by the horizontal, blue and red arrows in Fig.~\ref{f3}b). The rate of such a photo-assisted field emission is expected to depend non-linearly on the electric field strength and linearly on the laser intensity, as described by Fowler-Nordheim theory~\cite{photo-assisted-field-emission-1,photo-assisted-field-emission-2}. The predicted non-linear dependence on the electric field is observed in the current plotted in Fig.~\ref{f1}g  at higher gate voltages. The predicted linear dependence on the laser intensity is consistent with the laser power dependence shown in Fig.~\ref{f2}b (see also a previous study from some of us in Ref.~\cite{photo-induced-conductivity-2}), supporting that the brightening effect is indeed caused by a photo-assisted electroluminescence effect. 
Due to the continuous laser illumination, charge carriers get continuously injected into the defect donor and acceptor states, allowing them to follow the externally applied gate electric field by hopping between different defect states via Poole-Frenkel like emission~\cite{chiu_review_2014} (represented by the curved blue and red lines for donor and acceptor states, respectively; full circles represent electrons, hollow ones holes). Note that this excitation mechanism and the corresponding photo-induced transport of charges between hBN and adjacent 2-dimensional materials are in accordance with studies on photo-doping effects in van der Waals heterostructures using hBN as a dielectric layer~\cite{ju_photoinduced_2014,photo-induced-conductivity-4,photo-induced-conductivity-5,photo-induced-conductivity-3,photo-induced-conductivity-6}.
The injection of both electrons into the hBN from one gate electrode and holes into the hBN$^+$ from the other electrode (see Fig.~\ref{f3}b) leads to an explanation for the observed brightening of the emitter's ZPL and its connection to the leakage current across the hBN. The brightening effect can be understood in terms of photo-excited electron-hole pairs that recombine over the energy levels of the emitter and thus create additional electroluminescence - an effect that we denote as photo-assisted electroluminescence (see orange arrows in Fig.~\ref{f3}b). We support this proposed mechanism by theoretical simulations that can be found in the Supplemental Material I~\cite{SM}.  

\noindent Next we focus on the asymmetry of the brightening effect with respect to the gate voltage polarity and the resulting induced leakage current. As we will explain in the following, this asymmetry is a direct consequence of the asymmetry in the sample structure and can be described by theoretical simulations using a rate equation model (see Supplemental Material I~\cite{SM}). Due to the emitter generation process by annealing, it is likely that additional vacancy-complexes (e.g.~multi-vacancy or vacancy-substitutional defects) have been created in the hBN$^+$ that are not present in the untreated hBN layer. These defects may lack radiative recombination channels or be quenched due to their proximity to the FLG electrode~\cite{stewart_quantum_2021,PhysRevLett.132.196902} but they provide extra energy states within the hBN$^+$. Importantly, in contrast to carbon monomer defects present in both hBN layers - which create single donor or acceptor energy states - vacancy-complexes can exhibit energy levels with a series of occupied and unoccupied states extending over the entire band gap range at the location of the defect (see the thin stacked lines spanning over the hBN$^+$ band gap in Fig.~\ref{f3} and the more detailed discussion in the Supplemental Material G~\cite{SM})~\cite{PhysRevMaterials.1.071001,abdi_color_2018,energy-levels-1,energy-levels-2,energy-levels-3,energy-levels-4,energy-levels-5,energy-levels-6,energy-levels-7,sajid_defect_2018,abdi_color_2018,jara_first-principles_2021}. 
With this, the observed gate voltage asymmetry can now be explained under the assumption that the energy levels of these vacancy-complexes which are only present in the hBN$^+$ may allow for additional charge relaxation channels with differing relaxation rates for electrons and holes.
At negative gate voltages, holes from the TG are injected into the hBN layer and can reach the emitter in the hBN$^+$ by hopping between acceptor states (as for the positive gate voltage case). However, electrons injected from the BG now have additional relaxation channels within the hBN$^+$, making it less likely that they will reach the upper emitter level related to the optical emission.
Fig.~\ref{f3}c provides a simplified illustration of this process by showing that holes do not relax before reaching the emitter while electrons do (compare red lines in Fig.~\ref{f3}b to the blue lines in Fig.~\ref{f3}c). In fact, as we discuss in the Supplemental Material E~\cite{SM}, it is sufficient that only a fraction of holes does not relax before reaching the emitter. Provided that the relaxation of excited electrons into mid-gap states effectively prevents the occupation of the upper energy level of the emitter, the electroluminescence effect can be captured adequately by the presented model. 

%%%%%%%%%%%%%%%%%% Figure 4 %%%%%%%%%%%%%%%%%%%%%%
\begin{figure}[]
\centering
\includegraphics[width=\linewidth]{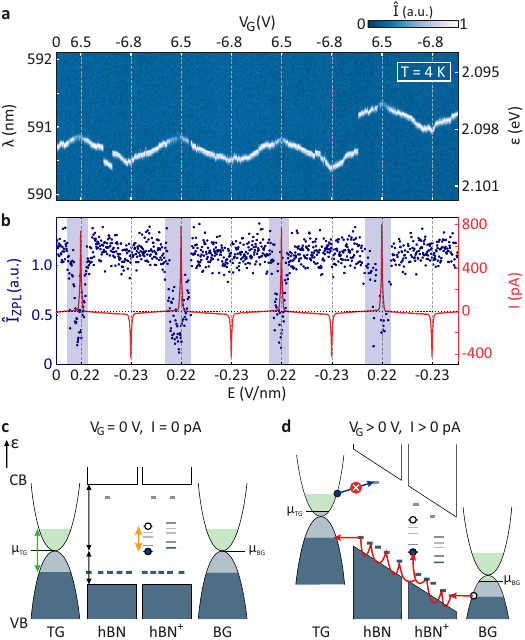}
\caption{
Gate voltage dependence of emitter intensity and leakage current of a second emitter in a different sample with a ZPL at 590.7\,nm measured at 4\,K with a laser power of $\mathrm{50\,\mu W}$.
(a) Emitter intensity as a function of the applied electric field. The emitter exhibits a reproducible linear Stark shift and voltage dependent changes in intensity.
(b) Integrated ZPL intensity of Lorentz fitted peaks from (a). The red curve depicts the simultaneously measured gate leakage current. The blue shaded areas mark the electric field ranges where the emitter intensity is decreased. As for the first emitter, a clear dependence of the intensity change on the gate leakage current is observed.
(c) and (d) Adapted schematic of the previously discussed model for this emitter. All notations like in Fig.~\ref{f3}. (c) depicts the case for $V_\mathrm{G} = 0$. The chemical potential of the FLG electrodes are shifted towards the hBN valence band corresponding to the case of p-doped hBN. (d) For $V_\mathrm{G} > 0$ electron injection and thus electron-hole recombination in the emitter levels is blocked. Occasional de-occupation of the lower emitter layer blocks its PL emission.}
\label{f4}
\end{figure}
%%%%%%%%%%%%%%%%%%%%%%%%%%%%%%%%%%%%%%%%%%%%%%%%%%%%%%

\noindent Interestingly, we also find hBN emitters which do not exhibit a brightening but rather show a current (gate voltage) induced \textit{de}crease in their emission intensity. Such an emitter is presented in Fig.~\ref{f4}. The emitter is located in a second sample with the same stacking order as shown in Fig.~\ref{f1}a. Both the position of the ZPL at 590.7\,nm and the strength of the linear Stark shift of $s = \mathrm{0.73 \pm 0.01\,(nm)/(V/nm)}$ are similar to the other emitter (compare Fig.~\ref{f1}f to Fig.~\ref{f4}a). However, at positive electric field values, the emitter intensity now decreases in contrast to the previously discussed increase in intensity. 
In analogy to the first emitter, the change in intensity (blue data points in Fig.~\ref{f4}b) coincides with the direction of the leakage current switching from negative to positive values (red line in Fig.~\ref{f4}b) \footnote{A positive leakage current is defined for electrons flowing from the top gate to the back gate}. 

\noindent We now demonstrate that our model allows us to understand this darkening effect by assuming different doping in the hBN layers of the two samples. 
For the first emitter, we assumed a rough balance of donor and acceptor states (Fig.~\ref{f3}), which resulted in the chemical potentials of the FLG electrodes being aligned approximately to the center of the band gap of the hBN layers. Such an alignment can indeed be observed in several angle-resolved photoemission spectroscopy (ARPES) or scanning tunneling spectroscopy (STS) studies~\cite{ARPES-intrinsic,energy-levels-3,energy-levels-1}. 
For the second emitter, we now assume p-doped hBN for which other studies have shown a pronounced shift of the chemical potentials towards the valence band of the hBN~\cite{ARPES-p-doped,Fermi-level-near-VB,ARPES-slightly-p-doped} (see Fig.~\ref{f4}c for $V_\mathrm{G} = 0$). This p-doping requires an excess of acceptor states near the valence band edge that leads to a shift in the band alignment of the hBN relative to the FLG electrodes, where the chemical potentials of the electrodes are now no longer in the center of the hBN band gap, but shifted closer to the hBN valence band edge. 
As a result, the dielectric strength of a structure with such a band alignment would be reduced as it is indeed observed for the second sample (see Supplemental Material F~\cite{SM}).
Based on this change in band alignment, the darkening of the emitter can now be understood from Fig.~\ref{f4}d. Here, the band alignment corresponds to p-doped hBN and is depicted for a positive gate voltage $V_\mathrm{G} > 0$, for which this emitter shows a decrease in intensity (in contrast to the brightening effect, compare Fig.~\ref{f1}f and Fig.~\ref{f3}b). 
Due to the shift of the chemical potential towards the valence band, the top gate can no longer inject electrons into donor states of the hBN capping layer (see red cross symbol in Fig.~\ref{f4}d). Instead, the whole leakage current over the hBN is carried via holes over states near the valence band edge (see curved red arrows). The lack of electron injection inhibits electron-hole pair recombination via the energy levels of the emitter, i.e.~there is no photo-assisted electroluminescence effect. Instead, the lower level of the emitter occasionally becomes unoccupied due to holes hopping between these states. During such times, the emitter can no longer be optically excited, leading to an effective decrease in emitter intensity.  

\textit{Conclusion}
In summary, we have demonstrated that the intensity of V-C type hBN emitters can be modulated by photo-assisted electroluminescence. For this, the necessary leakage current is injected into the hBN layers from few-layer graphene gates via photo-assisted field emission. Depending on the doping level of the hBN layers, we propose that the leakage current can either constitute of both electrons and holes or only of holes. The first case leads to the possibility of electron-hole pairs recombining at the location of the emitter, resulting in its brightening, whereas the second case results in the sporadic deoccupation of the emitter's lower energy level, resulting in the emitter getting even darker. Our findings highlight the necessity for a reliable benchmarking of hBN crystals~\cite{Ouaj_2023, Ouaj2024Sep} and the need for a careful determination of their doping levels in order to achieve electronic control over the brightness of dark quantum emitters in hBN.

\subsection*{Supplemental Material}
The Supplemental Material \cite{SM} contains 1.) details on sample fabrication, 2.) details on the measurement setup, 3.) measurements of emitter 1 at 4K, 4.) discussion on the magnitude of leakage current vs. increase in brightness of the emitter, 5.) a discussion on the contributions to leakage current, 6.) a discussion on the physical origin of the leakage current, 7.) a discussion on the electron-hole asymmetry, 8.) a discussion on the hysteresis effect of the gate-dependent brightening effect, and 9.) theoretical simulations of the charge distribution and relaxation processes., which includes Refs.\cite{taniguchi_synthesis_2007,uslu_open-source_2024,thermal-stability-hBN-1,thermal-stability-hBN-2,thureja2024electricallydefinedquantumdots,grzeszczyk_electroluminescence,carbon-domains-in-hBN,hBN-breakdown-1,hBN-breakdown-2,hBN-breakdown-3,hBN-breakdown-4,photo-assisted-field-emission-1,photo-assisted-field-emission-2,leakage-due-to-carbon-defects,defect-tunneling,jara_first-principles_2021,energy-levels-7,CBVN-center-for-quibit,carbon-contamination-1,carbon-contamination-2,carbon-contamination-3,mendelson_identifying_2021,mackoit-sinkeviciene_carbon_2019,energy-levels-3,energy-levels-2,energy-levels-6,carbon-dimer-energy-levels-1,PhysRevMaterials.1.071001,ju_photoinduced_2014,photo-induced-conductivity-2,photo-induced-conductivity-3,energy-levels-4,photo-induced-conductivity-4,photo-induced-conductivity-5,photo-induced-conductivity-6,photo-induced-conductivity-7,hBN-defects-hysteresis-noise,charge-trapping-sweep-rate-hBN}

\subsection*{Acknowledgments}
Funded by the Deutsche Forschungsgemeinschaft (DFG, German Research Foundation) under Germany's Excellence Strategy - Cluster of Excellence Matter and Light for Quantum Computing (ML4Q) EXC 2004/1 - 390534769 and by the Federal Ministry of Education and Research (BMBF) and the Ministry of Culture and Science of the German State of North Rhine-Westphalia (MKW) under the Excellence Strategy of the Federal Government and the Länder.
K.W. and T.T. acknowledge support from the JSPS KAKENHI (Grant Numbers 21H05233 and 23H02052) and World Premier International Research Center Initiative (WPI), MEXT, Japan.
This research was funded in part by the Austrian Science Fund (FWF) [10.55776/DOC142]. For open access purposes, the author has applied a CC BY public copyright license to any author accepted manuscript version arising from this submission.
Fabrication of the samples was supported by the Helmholtz Nano Facility (HNF) at the Forschungszentrum Jülich \cite{HNF}.

\subsection*{Competing interests}
The authors declare no competing interests.

\subsection*{Data availability}
The data supporting the findings of this study are available in a Zenodo repository under 10.5281/zenodo.14049616.

\end{document}

\title{Supplemental Material: Current-induced brightening of vacancy-related emitters in hexagonal boron nitride}

\author{Corinne Steiner}
\affiliation{JARA-FIT and 2nd Institute of Physics, RWTH Aachen University, 52074 Aachen, Germany}
\affiliation{Peter Gr\"unberg Institute  (PGI-9), Forschungszentrum J\"ulich, 52425 J\"ulich,~Germany}

\author{Rebecca Rahmel}
\affiliation{JARA-FIT and 2nd Institute of Physics, RWTH Aachen University, 52074 Aachen, Germany}

\author{Frank Volmer}
\affiliation{JARA-FIT and 2nd Institute of Physics, RWTH Aachen University, 52074 Aachen, Germany}
\affiliation{AMO GmbH, Advanced Microelectronic Center Aachen (AMICA), 52074, Aachen, Germany}

\author{Rika Windisch}
\affiliation{Institute for Theoretical Physics, TU Wien, 1040 Wien, Austria}

\author{Lars H. Janssen}
\affiliation{JARA-FIT and 2nd Institute of Physics, RWTH Aachen University, 52074 Aachen, Germany}

\author{Patricia Pesch}
\affiliation{JARA-FIT and 2nd Institute of Physics, RWTH Aachen University, 52074 Aachen, Germany}

\author{Kenji Watanabe}
\affiliation{Research Center for Electronic and Optical Materials, National Institute for Materials Science, 1-1 Namiki, Tsukuba 305-0044, Japan}

\author{Takashi Taniguchi}
\affiliation{Research Center for Materials Nanoarchitectonics, National Institute for Materials Science,  1-1 Namiki, Tsukuba 305-0044, Japan}

\author{Florian Libisch}
\affiliation{Institute for Theoretical Physics, TU Wien, 1040 Wien, Austria}

\author{Bernd Beschoten}
\affiliation{JARA-FIT and 2nd Institute of Physics, RWTH Aachen University, 52074 Aachen, Germany}

\author{Christoph Stampfer}
\affiliation{JARA-FIT and 2nd Institute of Physics, RWTH Aachen University, 52074 Aachen, Germany}
\affiliation{Peter Gr\"unberg Institute  (PGI-9), Forschungszentrum J\"ulich, 52425 J\"ulich,~Germany}

\author{Annika Kurzmann}
\affiliation{JARA-FIT and 2nd Institute of Physics, RWTH Aachen University, 52074 Aachen, Germany}
\affiliation{2nd Institute of Physics, University of Cologne, 50937 Köln, Germany}

\maketitle

\subsection{Sample fabrication}
Graphene (purchased from "NGS Naturgraphit") and hBN crystals grown by a high pressure high temperature method (NIMS \cite{taniguchi_synthesis_2007}) were exfoliated onto a $\mathrm{Si^{++}}$/$\mathrm{SiO_2}$ (90\,nm) substrate using a dicing tape (1007R Ultron). To identify suitable flakes for stacking, we use a home-built automatic flake search system \cite{uslu_open-source_2024}. The thicknesses of the selected flakes range from 10 to 50\,nm for hBN and from 3 to 4 layers for graphene. In order to create defects in the hBN, a subset of the exfoliated hBN flakes was subjected to annealing in an argon atmosphere for 30 minutes at 850°C, which is the temperature at which hBN starts to oxidize in ambient conditions \cite{thermal-stability-hBN-1,thermal-stability-hBN-2}. These flakes were used as emitter layers in the stacking process.

\noindent The assembly of the stacks is done via a dry van der Waals stacking method, using stamps consisting of polycarbonate (PC) and polydimethylsiloxane (PDMS). The pick-up temperatures were 110°C for graphene, 120°C for the annealed hBN and 90°C for the non-treated hBN. The stacking procedure was performed in two steps. First, the hBN emitter layer and the graphene back gate were assembled and placed onto a $\mathrm{Si^{++}}$/$\mathrm{SiO_2}$ (285\,nm) wafer with pre-patterned chromium/gold markers that are used for alignment in a following electron-beam lithography process. Subsequently, the wafer is baked at 165°C on a hotplate to ensure optimal adhesion between the stack and the substrate. The residual PC is then dissolved in chloroform, followed by additional cleaning of the substrate with acetone and isopropanol. In a second step, the graphene top gate and capping layer were assembled. To ensure a successful pick-up and a clean surface of the graphene top gate, an additional thin hBN layer was picked up first to prevent direct contact between the PC and the graphene. This upper half stack is then placed onto the previously assembled bottom half stack, followed by the same baking and cleaning steps described above.

\noindent The quality of each sample was evaluated through pre-characterization by atomic force microscopy (AFM) and room temperature Raman spectroscopy measurements to identify any contaminations, bubbles or broken flakes that may have been introduced during the stacking process. Additionally, the AFM measurements were used to determine the precise thickness of the hBN flakes. The electrical contacts for the top and back gate were fabricated using electron beam lithography to pattern the contact structures, followed by electron beam evaporation of 5\,nm Cr/70\,nm Au. 

\subsection{Measurement setup}
The samples were measured in an ATTO LIQUID 1000 cryostat at temperatures ranging from 4\,K to 150\,K. For the photoluminescence measurements we used a continuous-wave 532\,nm laser as excitation source. The laser is fiber-coupled to an optical head (attocube CFMI) where the light is directed and focused onto the sample by an objective (NA = 0.9, spot size of around 500\,nm). The emitted light is collected by the same objective, passed through an edge filter and coupled into a fiber to the spectrometer. The PL spectra of the emitters were obtained using gratings with 1200\,g/mm and 2400\,g/mm.

\noindent The gate voltage-dependent measurements were performed using a Keithley 2400 Source Measure Unit (SMU). This allows a voltage to be applied  to the gates while simultaneously measuring the gate leakage current.

\subsection{Characterisation of emitter 1 at 4K}
Additionally to the PL measurements at 77\,K, discussed in the main text, we performed 4\,K measurements at zero gate voltage of emitter 1 from the main text in order to characterize it with regard to its linewidth, time stability and laser power dependence. 

\begin{figure}[]
    \centering
    \includegraphics[width=\linewidth]{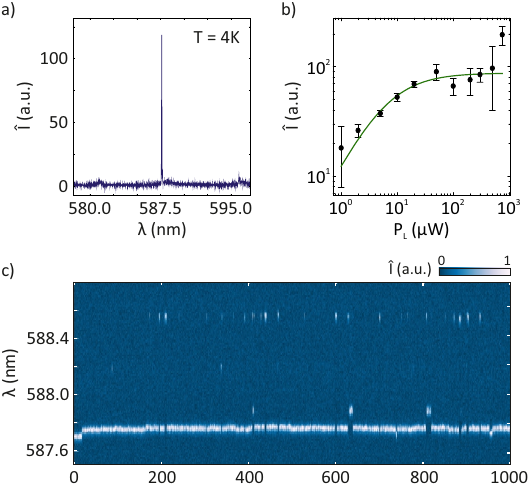}
    \caption{Emitter 1 from the main text measured at $\mathrm{T = 4\,K}$: Single PL spectra (a), laser power dependence (b) and PL spectra over time (c).}
    \label{S5_1}
\end{figure}

\noindent Fig.~\ref{S5_1} a) shows a PL spectra of this emitter at 4\,K that exhibits a single peak with very narrow linewidth $\approx \mathrm{0.1\,nm}$. This value is at the limit of the resolution of the spectrometer and indicates that the measured spectra corresponds to a single emitter source. We also observe the typical saturation of single photon emitters with increasing laser power as shown in Fig.~\ref{S5_1} b). By measuring successive spectra over time we are able to observe discrete jumps in the wavelength of the emitter which we attribute to charging events in the close vicinity of the emitter. A time series of emitter 1 is depicted in Fig.~\ref{S5_1} c). These spectral jumps are another indicator that the PL signal does not correspond to an emitter ensemble but indeed to a single emitter \cite{thureja2024electricallydefinedquantumdots}.\\

\subsection{Magnitude of leakage current vs. increase in brightness}
We note that even a tiny current of 1\,pA already corresponds to a flow of roughly six million electrons per second. On the other hand, the current-induced brightening effect only results in an increase of the emitter's intensity of about 100 counts per seconds. Accordingly, only a tiny fraction of the overall measured leakage current is actually responsible for the observed photo-assisted electroluminescence effect. 

\noindent There are two considerations that can explain the huge difference between the measured electron flow and the increase in the emitter's brightness: First, it should be noted that not all of the charges contributing to the measured leakage current will flow in close enough proximity to the emitter to be able to affect it. This will be discussed in detail in section\,\ref{Different contributions_of_the_leakage_current} of this  Supplemental Material. Secondly, in Fig.\,3b of the main manuscript we only depict the pathways of electrons and holes that result in the photo-assisted electroluminescence effect. However, next to electron-hole recombination over the energy states of the emitter, there are also many other pathways over which the injected electrons and holes can relax that are not shown in Fig.\,3b for sake of simplicity (like non-radiative relaxation over other defects or defect-to-band transitions that do not emit at well-defined energies and instead create broad background signals in photoluminescence measurements \cite{grzeszczyk_electroluminescence}).         

\begin{figure}[]
    \centering
    \includegraphics[width=\linewidth]{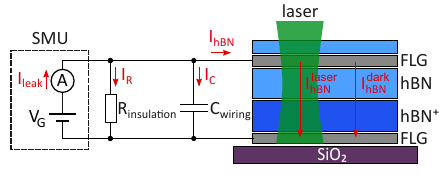}
    \caption{Schematic of both the measurement setup and the sample illustrating the different contributions of the overall measured leakage current.}
    \label{S1}
\end{figure}

\subsection{Different contributions of the leakage current}\label{Different contributions_of_the_leakage_current}
The measured leakage current $I_\textrm{leak}$ consists of three main contributions $I_\textrm{leak}=I_\textrm{hBN}+I_\textrm{R}+I_\textrm{C}$ as illustrated in Fig.\,\ref{S1} and discussed in the following.

\noindent $I_\textrm{hBN}$ is the current flowing over the hBN layers. This current can be divided in the current $I_\textrm{hBN}^\textrm{laser}$ that is flowing over the hBN at the position of the laser and the current $I_\textrm{hBN}^\textrm{dark}$ that is the sum of all charges flowing over the hBN at every other position. In case of spatially homogeneous hBN layers, it is safe to assume that the photo-assisted field emission results in $I_\textrm{hBN}^\textrm{laser} > I_\textrm{hBN}^\textrm{dark}$. As the laser is positioned at the location of the emitter, a significant part of $I_\textrm{hBN}$ will therefore flow in the vicinity of the emitter. However, as the laser spot size of around 500\,nm covers an area of the hBN that is quite large in comparison to the crystal defect that is constituting the emitter, many charge carriers might not flow in close enough proximity to the emitter to be able to interact with it.   
    
\noindent Furthermore, there may be cases where a significant portion of the current could actually flow in the non-illuminated parts of the sample, e.g.\ in hBN layers containing carbon-rich domains \cite{carbon-domains-in-hBN}. This would be the case if the emitter and therefore the laser spot are located in a part of the sample with a negligible concentration of carbon-related defects. In that case, both the effect of photo-assisted field emission over the high-quality parts of the hBN or the purely electric field driven emission in non-illuminated carbon rich domains can be the dominant contribution.    
    
\noindent The next contribution to the total measured leakage current is driven by the finite insulation resistance $R_\textrm{insulation}$ of the measurement setup (see Fig.\,\ref{S1}). $R_\textrm{insulation}$ in turn consists of three main contributions: 1.) the insulation resistances of the used wires, 2.) the insulation resistances of the mechanical switches in the breakout box (connector box) of the cryostat, and 3.) the insulation resistance between the two outputs of the SMU. The total resistance $R_\textrm{insulation}$ in the used setup normally lies in the range of several hundreds of G$\Omega$ to a few T$\Omega$. The exact value can vary with humidity levels in the laboratory, as small water films especially in the mechanical switches of the connector box or the woven loom cables (ribbon cables) used in the wiring of the cryostat are often the limiting factor of the insulation resistance. The corresponding current $I_\textrm{R}=V_\textrm{G}/R_\textrm{insulation}$ depends linearly on the applied gate voltage and is therefore partly responsible for the linear increase of the leakage current seen around $V_\textrm{G}=0\,V$ in Figs.\,1g and 4b of the main manuscript. 
    
\noindent The third contribution to the measured leakage current is the (dis-)charging current $I_\textrm{C}$ of the wires' capacitances $C_\textrm{wiring}$ to ground and is given as $I_\textrm{C}=C_\textrm{wiring}\cdot dV_\textrm{G}/dt $. As we ramp the gate voltage in a linear manner (i.e.\ $dV_\textrm{G}/dt=\textrm{const}$), the resulting current $I_\textrm{C}$ will also be constant and therefore creates an offset in the current measurements that are shown in Figs.\,1g and 4b of the main manuscript. Both the constant offset caused by $I_\textrm{C}$ and the linear contribution of $I_\textrm{R}$ can explain why the gate voltage range at which the brightening of the emitter occurs (shaded areas in Fig.\,1g of the main manuscript) do not perfectly coincidence with the zero crossings of the measured leakage current (crossings of the red line with the dotted line in Fig.\,1g of the main manuscript). Similarly, the uncertainty caused by $I_\textrm{C}$ and $I_\textrm{R}$ can also explain why the emitter sometimes seems to brighten even at small negative current values in Fig.\,1d of the main manuscript. 

\begin{figure*}[]
    \centering
    \includegraphics[width=\linewidth]{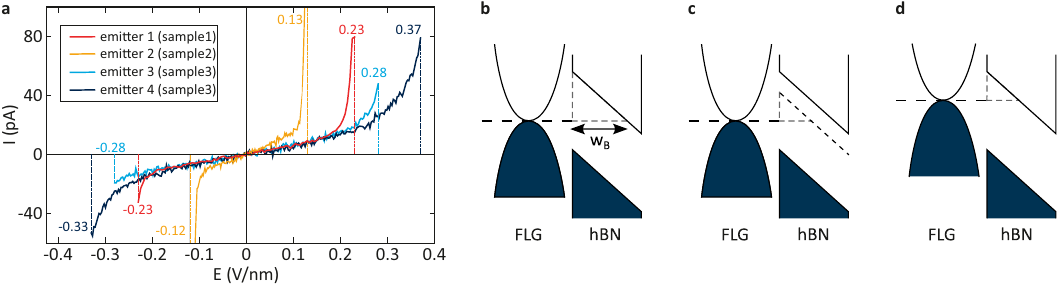}
    \caption{(a) Leakage current as a function of the electric field strength for four different emitters in three samples. The measurements were conducted under laser illumination at positions of emitters with laser powers of $\mathrm{30-50\mu W}$ at $\mathrm{T = 4\,K}$ (emitter 2-4) and $\mathrm{T = 77\,K}$ (emitter 1). (b) The highest possible dielectric strength is achieved when the hBN does not have any defect states and when the chemical potential of the gate electrode lies in the middle of the band gap. (c) Introducing defect states or (d) shifting the chemical potential of the gate electrode towards the conduction or valence band of the hBN will decrease the width $w_\textrm{B}$ of the tunnel barrier (see dashed lines in (b) to (d)) that must be overcome by the field emission.}
    \label{S2}
\end{figure*}

\subsection{Field-emission into defect states vs. band states}
\label{Field-emission}

The leakage current of sample 1 measured at the position of emitter 1 that shows the brightening effect undergoes a strong non-linear increase at electric field strengths that are rather small compared to reference emitter positions in other samples where we do not observe a current-induced change in emitter intensity. The respective leakage currents as a function of the electric field strength are depicted in Fig.~\ref{S2}a for emitter 1 that shows the brightening effect (red curve, sample 1 of Fig.\,1 in the main manuscript), for emitter position 2 that shows the current-induced decrease in emitter intensity (orange curve, sample 2 of Fig.\,4 in the main manuscript), and for two reference emitters in sample 3 that do not show any current-dependent changes in emitter intensity (dark and light blue curves). Note that, for Fig.\,\ref{S2}a we calculated the electric field strength not by the Lorentz local field approximation as in the main manuscript but by the simple plate capacitor model, i.e.\,$E=V_\text{G}/d$ with the gate voltage $V_\text{G}$ and the total thickness $d$ of both hBN emitter and capping layers. This approach guarantees better comparability to studies that have investigated the dielectric strength and breakdown voltages of hBN \cite{hBN-breakdown-1,hBN-breakdown-2,hBN-breakdown-3,hBN-breakdown-4} and that use the plate capacitor model. These studies report the onset of pronounced leakage currents across hBN at electric field ranges from 0.6\,V/nm to 1.0\,V/nm, which is larger than the onsets observed in our samples. Especially the ones with emitters 1 and 2 show low dielectric strengths of 0.23\,V/nm and 0.13\,V/nm, respectively. 

\noindent To reach dielectric strengths greater than 0.6\,V/nm we expect three main conditions to be required, depicted in Fig.\,\ref{S2}b: 1.) a negligible amount of defect states in the hBN layer, 2.) an alignment of the chemical potential of the gate electrode to the middle of the band gap, and 3.) the absence of light that would otherwise enable photo-assisted field-emission. Any deviation from these requirements would reduce the dielectric strength, as shown in Fig.~\ref{S2}c and \ref{S2}d for the introduction of defect states (c) or a shift of the chemical potential of the gate electrode (d). In both cases the width $w_\textrm{B}$ of the tunnel barrier that must be overcome by the field emission is diminished (see gray dashed lines). 

\noindent We therefore propose that emitter 4 (dark blue curve in Fig.\,\ref{S2}a) comes closest to the condition that is depicted in Fig.\,\ref{S2}b and that it is mainly photo-induced emission that reduces the dielectric strength of the position of this emitter to values around 0.37\,V/nm. For the positions of emitter 3 and especially 1, the presence of a larger density of donor and acceptor states reduces the dielectric strength further. Finally, the lowest dielectric strength observed at the position of emitter 2 can be explained by an additional shift of the chemical potentials of the gate electrodes. This shift is in complete accordance to the model that is explaining the current-induced decrease in its intensity shown in Figs.\,4c and 4d of the main manuscript. 

\noindent Overall, the reduced dielectric strengths of samples 1 and 2 can be readily explained by photo-assisted field emission \cite{photo-assisted-field-emission-1,photo-assisted-field-emission-2} into carbon-related defect states present in hBN \cite{leakage-due-to-carbon-defects,defect-tunneling}. The absence of any current-induced changes in the intensities of emitter 3 and 4 can accordingly be explained if we assume a negligible density of donor and acceptor states in sample 3. The charges are therefore not injected into defect states but photo-excited directly into band states, however at much higher electric field strengths. Once inside the bands of the hBN and with no defect states that can trap the charges, the charges can freely flow in the hBN and no longer require laser excitation to hop from one defect to the next (as for the Poole-Frenkel like emission depicted in Fig.\,3 of the main manuscript). Accordingly, the laser loses its ability to funnel the leakage current within the area of the sample in which the emitter is located. Consequently, any current-induced changes in emitter intensity are suppressed.

\begin{figure*}[]
    \centering
    \includegraphics[width=\linewidth]{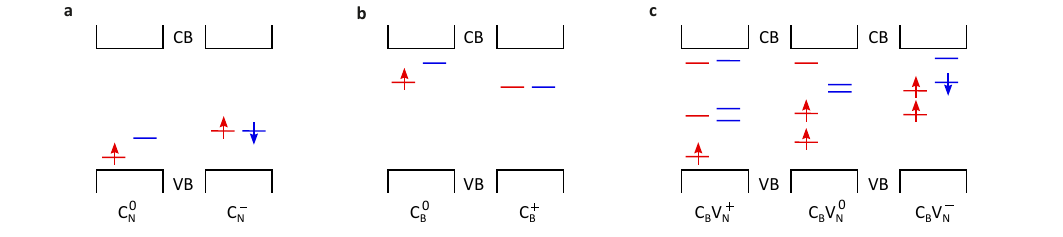}
    \caption{Simplified energy levels of (a) the acceptor-like C$_\textrm{N}$ defect, (b) the donor-like C$_\textrm{B}$ defect, and (c) the C$_\textrm{B}$V$_\textrm{N}$ defect reproduced from Refs.\,\cite{jara_first-principles_2021,energy-levels-7,CBVN-center-for-quibit}. Red and blue colors represent spin up and down states, respectively. Arrows symbolize the occupation of the energy levels by electrons. For each type of defect both the neutral state and the energetically allowed charged states are depicted. Additional energy levels of the defects that lie within either conduction band (CB) or valence band (VB) of the hBN were omitted.}
    \label{S3}
\end{figure*}

\subsection{Possible origin of the electron-hole asymmetry}
To explain the asymmetry of the brightening effect with respect to the direction of the gate voltage-induced leakage current, we have assumed in the main manuscript that  excited electrons and holes undergo different relaxation rates within the emitter layer. Although identifying the exact origin of these asymmetric rates is beyond the scope of our study, we would like to propose one possible explanation. For this we focus on the types of defects we can expect in both the emitter and capping layer and their energy levels. We start with carbon monomer defects, as carbon is one of the most ubiquitous contaminations in hBN \cite{carbon-contamination-1,carbon-contamination-2,carbon-contamination-3,mendelson_identifying_2021,mackoit-sinkeviciene_carbon_2019,carbon-domains-in-hBN,energy-levels-3,energy-levels-7} and therefore can be expected in both the emitter and the capping layer.

\noindent Depending on whether a carbon atom replaces a nitrogen (C$_\textrm{N}$) or boron atom (C$_\textrm{B}$), substitutional carbon defects can form acceptor or donor states, respectively \cite{energy-levels-2,energy-levels-3,energy-levels-7,energy-levels-6,mackoit-sinkeviciene_carbon_2019,carbon-dimer-energy-levels-1}. The energy states of the monomer defects are depicted in Figs.\,\ref{S3}a and \ref{S3}b for both the neutral (C$_\textrm{N}^{0}$, C$_\textrm{B}^{0}$) and the energetically allowed charged states (C$_\textrm{N}^{-}$, C$_\textrm{B}^{+}$) (reproduced after Refs.\,\cite{jara_first-principles_2021,energy-levels-7}). Red and blue colors represent spin up and down states, whereas arrows symbolize the occupation of the energy levels by electrons. As discussed in the main manuscript and in even more detail in section\,\ref{Field-emission}, these carbon-related defects are very likely to carry the leakage current over the hBN layers. However, they are no source of single photon emission, as they either do not possess bright transitions between defect states or the transitions exhibit such low oscillator strengths that they are inconsistent to an observable emitter \cite{jara_first-principles_2021,PhysRevMaterials.1.071001,mendelson_identifying_2021}.

\noindent The whole situation changes if we now consider vacancy-related defects created by the annealing process in the emitter layer. As an example, we discuss the dimer defect C$_\textrm{B}$V$_\textrm{N}$ that consists of a vacancy at a nitrogen site adjacent to a substitutional carbon defect at a boron site. The simplified energy levels are depicted in Fig.\,\ref{S3}c (reproduced after Ref.\,\cite{CBVN-center-for-quibit}, additional energy levels within either conduction or valence band of the hBN were omitted for the sake of simplicity). In contrast to the substitutional carbon defects, both the negatively and positively charged states of this defect are energetically allowed. Therefore, depending on the applied voltages and position of the Fermi level, the flow of either electrons or holes over these defects will momentarily charge them negatively or positively, respectively. However, as seen in Fig.\,\ref{S3}c, adding or removing one electron from the defect has a significant impact on the energies of all occupied and unoccupied states. We propose that within this many-particle picture the rate at which excited holes or electrons relax into mid-gap states of this emitter might differ, which in turn would explain the observed asymmetry. 

\begin{table}[h]
    \centering
    \caption{On- and off values of intensity switches}
    \begin{tabular}{c|c|c}
         Sweep No. & $\mathrm{V_{on}}$ & $\mathrm{V_{off}}$ \\
         \hline
         1 & $\mathrm{4.05 \pm 0.37}$ & $\mathrm{7.11 \pm 0.10}$\\
         2 & $\mathrm{2.51 \pm 0.18}$ & $\mathrm{6.48 \pm 0.06}$\\
         3 & $\mathrm{1.04 \pm 0.11}$ & $\mathrm{7.02 \pm 0.19}$\\
    \end{tabular}
    \label{t1}
\end{table}

\subsection{Drift and hysteresis of intensity variations}

By comparing the gate voltage values where the emitter of Fig.\,1 in the main manuscript switches its intensity from dark to bright ($V_\mathrm{on}$) with the values where the intensity switches back to its darker state ($V_\mathrm{off}$), two observations can be made (see table\,\ref{t1}): 1.) $V_\mathrm{on}$ drifts over time towards lower values, and 2.) $V_\mathrm{on}$ and $V_\mathrm{off}$ differ by several volts, meaning that there is a hysteresis in the intensity variations.

\noindent As we will discuss in the following, we attribute these two observations to time and illumination-dependent charging effects of defects in hBN (see also theoretical calculations in Section \ref{SI_theory}). Such charging effects have been known for quite some time under the name of photo-doping effect \cite{ju_photoinduced_2014,photo-induced-conductivity-2,photo-induced-conductivity-3,energy-levels-4,photo-induced-conductivity-4,photo-induced-conductivity-5,photo-induced-conductivity-6}. The charges that were put on defect sites in the hBN through photo-assisted field-emission can exhibit retention times within the hBN layer between seconds and days once the laser is switched off \cite{ju_photoinduced_2014,photo-induced-conductivity-4,photo-induced-conductivity-5,photo-induced-conductivity-6,photo-induced-conductivity-7}.
Without illumination trapping and detrapping of charged defects have been measured in the range of around 100\,ms to 10\,seconds depending on temperature and gate voltage \cite{leakage-due-to-carbon-defects}. Overall, charged defects within the hBN layer are generally assumed to be the reason for hysteresis effects that are measurable on timescales from milliseconds to minutes \cite{hBN-defects-hysteresis-noise,charge-trapping-sweep-rate-hBN}.

\subsection{Theoretical simulations}
\label{SI_theory}
To elucidate the charge distribution and relaxation processes, we model the dynamics of the capping (hBN) and emitter (hBN$^+$) layer stack using a set of coupled rate-equations describing the transitions between the different inter-band states within the hBN. We choose $z$ as the direction perpendicular to the hBN surface (thus treating an effective 1D problem given the in-plane periodicity. The occupations $n_j(z)$ of the different defect levels $j$ within the sample are updated each time step according to the different transition and hopping probabilities, including effects such as Pauli blocking once defect states fill up. Simultaneously, the resulting charge distribution is included
in a 1D Poisson equation to determine the local electric fields within the sample, which drives the charge relaxation processes.
Within hBN$^+$, we include 5 different defect levels $L_j$, $j=1\ldots 5$, with $L_1$ the acceptor and $L_5$ the donor states related to the carbon monomer defects, with an energetic separation $\varepsilon_5 - \varepsilon_1 = 5.5$\,eV. Three intermediate levels $L_2,\ L_3,\ L_4$ energetically in between ($\varepsilon(L_j) < \varepsilon(L_{j+1}$) represent localized, e.g., vacancy-related, defect states induced thermally only in the hBN$^+$ layer. Optical transitions may occur at one defect close to the hBN/hBN$^+$ interface between $L_2$ and $L_4$ ($\varepsilon_4-\varepsilon_2 = 2.1$\,eV), with the intermediate level $L_3$ representing all additional energy levels of vacancy-related defects that do not contribute to the emission of the discussed emitter.

\noindent To model the electrostatics due to the occupation of defect levels, we solve Poisson's equation after each time step, using the current charge distribution as well as the appropriate boundary conditions due to the applied electrostatic potential of the graphene layers at $z=0$ and $z=W$. We thus obtain a local electric field $F(z)$ within the stack, which in turn influences the motion of the charge carriers. 
Transitions between defect levels are described by decay constants $\gamma_{L_{j+1} \rightarrow L_{j}}$, while hopping between adjacent defect levels of the same energy happens with probability $\gamma^{\text{hop}}_{L_j}$. For simplicity, we do not include further transitions such as
$\gamma_{L_{j+2} \rightarrow L_{j}}$, except for the direct optical transition $\gamma^{\text{opt}}_{L_{j+2} \rightarrow L_{j}}$ at the interface.
The rate equations for each point $z_i$ in the emitter layer (excluding for brevity the optical transitions) are then \\
\begin{align*}
    \partial_t n_{L_j}(z_i) &= \gamma_{{L_{j+1}}\rightarrow L_j} \cdot n_{j+1}(z_i) \cdot (1-n_{L_j}(z_i)) 
    \\& - \gamma_{L_j\rightarrow L_{j-1}} \cdot n_{L_j}(z_i) 
    \cdot (1-n_{L_{j-1}}(z_i))
    \\& - \gamma^{hop}_{L_j} \cdot F(z_{i + \frac{1}{2}}) \cdot n_{L_j}(z_i) \cdot (1-n_{L_j}(z_{i+1}))
    \\& + \gamma^{hop}_{L_j} \cdot F(z_{i - \frac{1}{2}}) \cdot n_{L_j}(z_{i-1}) \cdot (1-n_{L_j}(z_i)).
\end{align*}
In the capping layer the equations are essentially the same, but without the transitions to and from vacancy-related defects.

\begin{figure}[H]
    \centering
    \includegraphics[width=\linewidth]{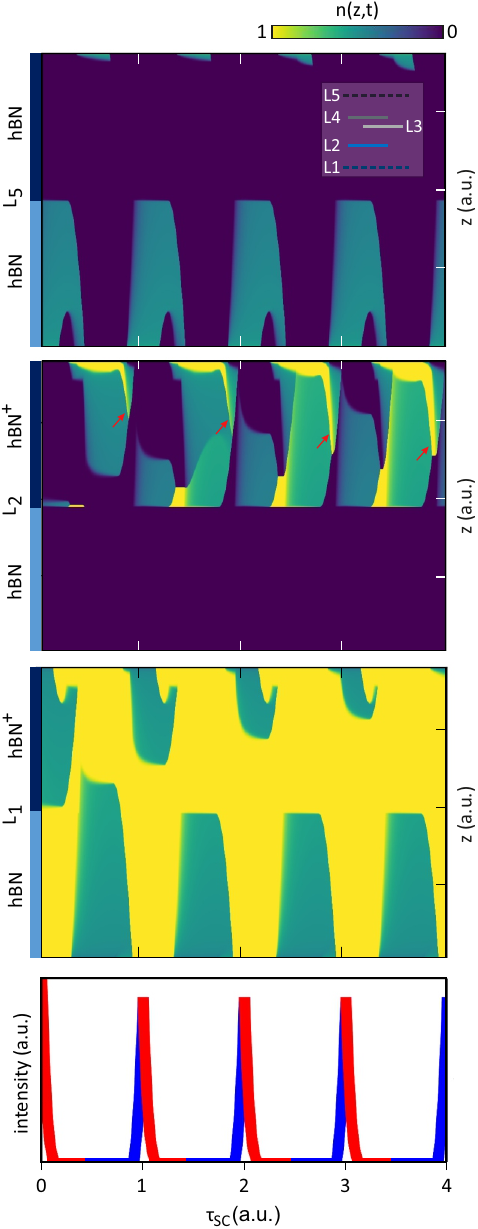}
    \caption{Simulation of the evolving occupation of inter-gap states $L_5$, $L_2$ and $L_1$ as a function of time and position within the hBN/hBN$^+$ stack (see labels), as well as of the optical intensity over time. Time axis is the same in all plots. Intensity plot is colored blue (red) when gate voltage is increasing (decreasing) at top graphene layer.} 
    \label{S4}
\end{figure}

\noindent We now perform a number of simulation cycles $N_{cycles}$ in which we propagate the rate equations, calculate the new occupations, and from these solve Poisson's equation to obtain the updated Electric Potential at each point $z_i$, due to the transitions. We also sweep the chemical potential of the FLG ``leads'' with the same triangular pattern as in the experiment with a sweep-cycle length $\tau_{SC}$ and define the timescales for the individual processes according to $\tau_{\mathrm{SC}}$. The rate equations are evaluated after a time step of $\Delta t=3 \cdot 10^{-6} \tau_{SC}$.
We find optical activity only when holes entering the stack from the capping-layer side ($V_{\mathrm{G}}>0$ in the experiment) de-excite optically at the hBN-hBN$^+$ interface (see Fig.~\ref{S4}), in line with the experimental observations (see Fig.~1\,f). In particular, electronic de-excitation from $L_5$ into lower-lying states available in the emitter layer happens too fast for electrons to propagate all the way to the interface when entering the stack from the emitter side (see leftmost plot in Fig.~\ref{S4}), corroborating the model discussed in the main text. As the voltage becomes positive again, the electrons start to leave the emitter layer (see $L_5$ in Fig.~\ref{S4}). In Level $L_2$ however, they temporarily accumulate at the boundary of the emitter layer, leading to a sharp peak in the occupation plot, until they leave on the emitter-layer side or recombine with available holes in Level $L_1$. This leads to the surprising observation of the bright feature in $L_2$ apparently moving against the direction of the applied voltage (see red arrows in Fig.~\ref{S4}). Akin to a shock wave in fluid dynamics or a traffic wave in a line of moving cars, electrons need to break because the hopping $L_2$ states in front of them are occupied, leading to a disturbance that briefly moves in the opposite direction from the electrons. A similar phenomenon occurs when the voltage switches to a negative value, as the remaining electrons in Level $L_2$ accumulate near the emitter/capping layer interface until the holes in Level $L_1$ reach the interface and recombination occurs.
\newpage
